\documentclass[]{nature}
\makeatletter\if@twocolumn\PassOptionsToPackage{switch}{lineno}\else\fi\makeatother

\usepackage{tabulary,graphicx,amsmath,amsfonts,amssymb}
\usepackage[utf8x]{inputenc}
\usepackage{amsmath}
\usepackage{graphicx}
\usepackage{color}
\usepackage[10pt]{moresize}
\usepackage{amsfonts}
\usepackage{amssymb}
\usepackage{amscd}
\usepackage{enumerate}
\usepackage{epsfig}
\usepackage{subfigure}
\usepackage{graphicx}
\usepackage{bm}
\usepackage{color}
\usepackage{epstopdf}
\usepackage{physics}

\makeatletter
\renewenvironment{figure}
               {\@float{figure}}
               {\end@float}
\renewenvironment{figure*}
               {\@dblfloat{figure}}
               {\end@dblfloat}
\renewenvironment{table}
               {\@float{table}}
               {\end@float}
\renewenvironment{table*}
               {\@dblfloat{table}}
               {\end@dblfloat}

\makeatother

\usepackage{url,multirow,morefloats,floatflt,cancel,tfrupee}
\makeatletter

\AtBeginDocument{\@ifpackageloaded{textcomp}{}{\usepackage{textcomp}}}
\makeatother
\usepackage{colortbl}
\usepackage{xcolor}
\usepackage{pifont}
\usepackage[nointegrals]{wasysym}
\makeatletter

\def\mcWidth#1{\csname TY@F#1\endcsname+\tabcolsep}

\def\cAlignHack{\rightskip\@flushglue\leftskip\@flushglue\parindent\z@\parfillskip\z@skip}
\def\rAlignHack{\rightskip\z@skip\leftskip\@flushglue \parindent\z@\parfillskip\z@skip}

\usepackage{ifxetex}
\ifxetex\else\if@twocolumn\usepackage{dblfloatfix}\fi\fi

\AtBeginDocument{
\expandafter\ifx\csname eqalign\endcsname\relax
\def\eqalign#1{\null\vcenter{\def\\{\cr}\openup\jot\m@th
  \ialign{\strut$\displaystyle{##}$\hfil&$\displaystyle{{}##}$\hfil
      \crcr#1\crcr}}\,}
\fi
}

\AtBeginDocument{%
  \@ifpackageloaded{endfloat}%
   {\renewcommand\efloat@iwrite[1]{\immediate\expandafter\protected@write\csname efloat@post#1\endcsname{}}}{}%
}%

\def\BreakURLText#1{\@tfor\brk@tempa:=#1\do{\brk@tempa\hskip0pt}}
\let\lt=<
\let\gt=>
\def\processVert{\ifmmode|\else\textbar\fi}

\@ifundefined{subparagraph}{
\def\subparagraph{\@startsection{paragraph}{5}{2\parindent}{0ex plus 0.1ex minus 0.1ex}%
{0ex}{\normalfont\small\itshape}}%
}{}

\newcommand\role[1]{\unskip}
\newcommand\aucollab[1]{\unskip}

\@ifundefined{tsGraphicsScaleX}{\gdef\tsGraphicsScaleX{1}}{}
\@ifundefined{tsGraphicsScaleY}{\gdef\tsGraphicsScaleY{.9}}{}
\def\checkGraphicsWidth{\ifdim\Gin@nat@width>\linewidth
	\tsGraphicsScaleX\linewidth\else\Gin@nat@width\fi}

\def\checkGraphicsHeight{\ifdim\Gin@nat@height>.9\textheight
	\tsGraphicsScaleY\textheight\else\Gin@nat@height\fi}

\def\fixFloatSize#1{}
\let\ts@includegraphics\includegraphics

\def\inlinegraphic[#1]#2{{\edef\@tempa{#1}\edef\baseline@shift{\ifx\@tempa\@empty0\else#1\fi}\edef\tempZ{\the\numexpr(\numexpr(\baseline@shift*\f@size/100))}\protect\raisebox{\tempZ pt}{\ts@includegraphics{#2}}}}

\AtBeginDocument{\def\includegraphics{\@ifnextchar[{\ts@includegraphics}{\ts@includegraphics[width=\checkGraphicsWidth,height=\checkGraphicsHeight,keepaspectratio]}}}

\DeclareMathAlphabet{\mathpzc}{OT1}{pzc}{m}{it}

\def\URL#1#2{\@ifundefined{href}{#2}{\href{#1}{#2}}}

\def\UrlOrds{\do\*\do\-\do\~\do\'\do\"\do\-}%
\g@addto@macro{\UrlBreaks}{\UrlOrds}

\@ifundefined{quoteAttrib}
	{}
	{}

\@ifundefined{titlequoteAttrib}
	{}{}

\newenvironment{title-quote}
	{\list{}{\fontsize{10pt}{12pt}\selectfont\leftmargin.5in\itshape\rightmargin\leftmargin}%
  \item\relax}
  {\endlist}

\makeatother

\usepackage[nomarkers,tablesfirst,nolists]{endfloat}

\def\fixFloatSize#1{}
\begin{document}

\title{Quantum Random Number Generation with Uncharacterized Laser and Sunlight}

\author{Yu-Huai Li$^{1,2}$, Xuan Han$^{1,2}$, Yuan Cao$^{1,2}$, Xiao Yuan$^{1,2,3}$, Zheng-Ping Li$^{1,2}$, Jian-Yu Guan$^{1,2}$, Juan Yin$^{1,2}$, Qiang Zhang$^{1,2}$, Xiongfeng Ma$^{3}$, Cheng-Zhi Peng$^{1,2}$, Jian-Wei Pan$^{1,2}$}

\maketitle

\begin{affiliations}
\item Shanghai Branch, National Laboratory for Physical Sciences at Microscale and Department of Modern Physics, University of Science and Technology of China, Shanghai 201315, China.
\item Synergetic Innovation Center of Quantum Information and Quantum Physics, University of Science and Technology of China, Hefei, Anhui 230026, China.
\item Center for Quantum Information, Institute for Interdisciplinary Information Sciences, Tsinghua University, Beijing 100084, China.
\\
\end{affiliations}

\begin{abstract}
The entropy or randomness source is an essential ingredient in random number generation. Quantum random number generators generally require well modeled and calibrated light sources, such as a laser, to generate randomness. With uncharacterized light sources, such as sunlight or an uncharacterized laser, genuine randomness is practically hard to be quantified or extracted owing to its unknown or complicated structure. By exploiting a recently proposed source-independent randomness generation protocol, we theoretically modify it by considering practical issues and experimentally realize the modified scheme with an uncharacterized laser and a sunlight source. The extracted randomness is guaranteed to be secure independent of its source and the randomness generation speed reaches 1 Mbps, three orders of magnitude higher than the original realization. Our result signifies the power of quantum technology in randomness generation and paves the way to high-speed semi-self-testing quantum random number generators with practical light sources.
\end{abstract}

\section*{Introduction}
Random numbers play a vital role in various tasks, such as cryptography,\cite{Shannon49} numerical simulation,\cite{Metropolis49} and lottery. For example, in the well-known quantum key distribution (QKD) protocol proposed by Bennett and Brassard,\cite{BB84} the security is guaranteed by random choices of the encoding and measurement bases. Distinct from deterministic evolution of classical processes, quantum mechanics endows the capability of generating genuine randomness by collapsing the coherence in the measurement basis.\cite{YuanPhysRevA2015,yuan2016interplay}

According to the generation speed and the randomness reliability, quantum random number generators (QRNGs) can be categorized into the following three types. Practical QRNGs, which assume well characterized devices, normally have a fast generation speed.\cite{YLD2014,AAJ2014,nie201568} Fully self-testing QRNGs, which adopt no assumptions on device implementations, generally have a low randomness generation speed owing to the stringent requirements.\cite{Pironio10,giustina2013bell,Liu2018,Bierhorst2018,PhysRevLett.120.010503} Semi-self-testing QRNGs lies somewhere in between, which have certain assumptions of device implementations while have high randomness generation speed in the meantime.\cite{Lunghi15,Cao16,2015arXiv150907390M,Avesani2018,PhysRevApplied.7.054018,Z1,Xu_2019,A2} We refer to Ref.~\cite{ma2015quantum} for a detailed review of the developments of different types of QRNGs. For these three types, a trade-off between the randomness generation speed and the randomness reliability exists in practice. In many tasks such as QKD, both the randomness generation speed and the reliability are required in order to ensure the key generation rate and the security. For those tasks, semi-self-testing QRNGs serve as promising candidates that fulfill both requirements.

Recently, several semi-self-testing QRNG schemes have been proposed.\cite{Lunghi15,Cao16,2015arXiv150907390M,Avesani2018,PhysRevApplied.7.054018,Z1,Xu_2019,A2} By assuming the underlying dimension and the independence of the source and the measurement, a QRNG scheme\cite{Lunghi15} has been proposed such that the output randomness can be self-tested. While, as the randomness is certified by the input and output statistics, the random number generation rate is only about 23 bps. The generation rate was further improved to the order of MHz with input and output statistics and weaker assumptions.\cite{PhysRevApplied.7.054018,Z1} Later, with trusted measurement but uncharacterized randomness source, a source-independent (SI) QRNG scheme is proposed,\cite{Cao16} where the randomness generation speed is analyzed to be comparable to practical QRNGs that has characterized devices. Conventionally, QRNGs make use of special light sources, such as lasers, and specific physical model to characterize the randomness source. With more common light sources, such as sunlight, and no assumptions of the randomness origin, the SI-QRNG scheme can still faithfully generate random numbers.

In this work, we explore randomness generation with general light sources, laser and sunlight without assuming the physical structures. By exploiting the SI-QRNG scheme\cite{Cao16} and considering practical issues of measurement device imperfections, we experimentally demonstrate the possibility of fast and reliable randomness generation.

\section*{Results}

\subsection{Scheme}

\begin{figure}[htbp]
\centering\includegraphics[width=8cm]{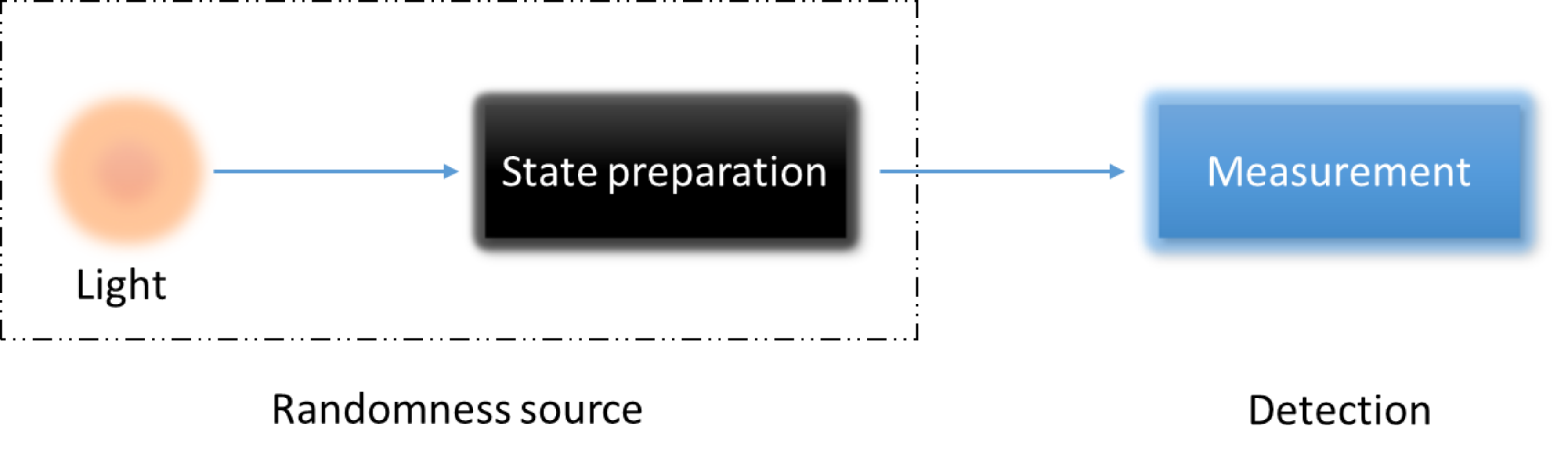}
\caption{Source independent randomness generation with uncharacterized light source and state preparation.}
\label{fig:scheme}
\end{figure}

As shown in Fig.~\ref{fig:scheme}, a conventional QRNG is composed of the randomness source and the detection device.\cite{ma2015quantum} In the source part, it consists of a light source and a state preparation device. Generally, practical QRNGs\cite{YLD2014,AAJ2014,nie201568} make use of specific models to describe the structure of the randomness source. While the SI-QRNG scheme\cite{Cao16} supplies the possibility of randomness generation without assuming neither the light source nor the state preparation devices.

First, we review the concept of SI-QRNG based on the scheme of Ref.~\cite{Cao16}. The source independent randomness generation procedure is summarized by source, squashing, random sampling, parameter estimation, and randomness extraction, as follows.
\begin{itemize}
  \item The state preparation device is expected to prepare the light in the polarization state of $\ket{+} = (\ket{H}+\ket{V})/\sqrt{2}$, where $Z = \{\ket{H},\ket{V}\}$ is the computational basis. While no assumption is made on the photon source, it is untrusted and could be controlled by Eve. Thus, the actual prepared quantum state may have an arbitrary and unknown dimension. Randomness can still be quantified and extracted with the following steps.

  \item The squashing process maps arbitrary quantum states into qubits and vacuum states. The vacuum components are regarded as loss. In practice, the squashing process can be realized by adding a series of spectrum, spatial and temporal filters to post-select the expected optical modes.

  \item In the measurement, we randomly choose the $X = \{\ket{\pm} = (\ket{H}\pm\ket{V})/\sqrt{2}\}$ and $Z$ basis to measure according to a random seed. The random seed needed is exponentially smaller than the number of extracted random bits.\cite{Cao16}

  Suppose that the number of runs in total is $N$, including $N_X$ in the $X$ basis and $N_Z$ in the $Z$ basis. Due to the loss, the output of detection could be null. In this case, the number of qubits measured in total is $n$, including $n_X$ in the $X$ basis and $n_Z$ in the $Z$ basis. It is worth noting that this protocol is loss tolerance. In the ideal case, the measurement device chooses the measurement basis after confirming the received state is not a vacuum state. In practice, the measurement basis is chosen before the confirmation of loss, which is usually done by observing whether detectors click or not. However, the detection device does not anticipate the position of losses. Thus the effect of loss only decreases $n_X$ and $n_Z$, but the positions of the effective X and Z measurements are still uniformly random.

  \item When measuring in the $X$ basis, the result of $\ket{-}$ is defined to be an error, and a double click is considered as a half error. Then, we can evaluate the phase error $e_{pZ}$ in the $Z$ basis according to the bit error rate $e_{bX}$ in the X basis and its statistical deviation $\theta$\cite{PhysRevA.81.012318} according to
      \begin{equation} \label{eq:Ptheta}
\begin{aligned}
\varepsilon_\theta &= \text{Prob}(e_{pZ}>e_{bX}+\theta)  \\
&\le \frac{1}{\sqrt{q_X(1-q_X)e_{bX}(1-e_{bX})n}}2^{-n\xi(\theta)},
\end{aligned}
\end{equation}
where $\xi(\theta)= H(e_{bX}+\theta-q_X\theta)-q_X H(e_{bX})-(1-q_X)H(e_{bX}+\theta)$. Here $q_X = n_X/n$ is the ratio of the $X$ basis measurement. $H(x) = -x\log_2(x)-(1-x)\log_2(1-x)$ is the binary Shannon entropy function.

Since the dimension of the source is unlimited, it may emit multiphoton states. When using threshold detectors, multiphoton states may cause double clicks, which directly contribute to the error rate $e_{bX}$ and decrease the number of extracted random bits.

\item By utilizing the Toeplitz-matrix hashing method,\cite{MANSOUR1993121} the phase error can be corrected by consuming $n_ZH(e_{bX}+\theta)$ number of bits with a failing probability of $2^{-t_e}$.\cite{Ma2011Finite} Thus, we can finally extract
\begin{equation}\label{EQ_R}
  R_0 = n_Z-n_ZH(e_{bX}+\theta)-t_e
\end{equation}
number of random bits and the final failure probability (in trace-distance measure) is given by $\varepsilon=\sqrt{(\varepsilon_\theta+2^{-t_e})(2-\varepsilon_\theta-2^{-t_e})}$.
\end{itemize}

Here, $R_0$ is the number of extracted random bits without considering the imperfections of the measurement devices. In practice, the measurement bases may not be exactly complementary to each other, and the detection efficiencies of the two detectors might be different. With a slight modification to Eq. (2), the number of extracted random bits with imperfect measurement devices $R_{final}$ can still be quantified when these imperfections are characterized. On the other hand, dark counts of the detectors may also increase the phase error rate and decrease the number of extracted random bits. Since the dark counts are independent with respect to the measurement basis, the effect of dark counts can be regarded as noise of the photon source which has already been considered in the analysis.

In the original theoretical proposal, the $X$ and $Z$ basis measurements are assumed perfect. In our work, we also take measurement imperfections into account. Specifically, we consider the case that the actual measurement bases $X^\prime$ and $Z^\prime$ are not complementary to each other. In this case, we can make use of the general uncertainty relation for two general bases,\cite{RevModPhys.89.015002}
\begin{equation}
	H(Z')\ge -\log_2{\max_{x^\prime,z^\prime}{|\langle x^\prime\ket{z^\prime}|^2}}-H(X'),
\end{equation}
where $\{\ket{x^\prime}\}$ and $\{\ket{z^\prime}\}$ are respectively the eigenstates of $X^\prime$ and $Z^\prime$, and $H(Z')$ and $H(X')$ are respectively the entropy of the measurement outcome of $X^\prime$ and $Z^\prime$. In quantum random number generation, we can regard $-\log_2{\max_{x^\prime,z^\prime}{|\langle x^\prime\ket{z^\prime}|^2}}$ as the randomness that we can obtained by measuring an eigenstate of the $X'$ basis, and regard $H(X')$ as the amount of states that are required to distill the eigenstate \cite{PhysRevLett.116.120404,yuan2016interplay}. That is, given $N$ copies of the quantum state $\rho$, one can effectively first perform a dephasing operation in the $X'$ basis to collapse them into one of its eigenstates. Then we aim to distill the dephased state into a specific eigenstate, which costs $NH(X')$ copies of states. For each eigenstate, it generates $-\log_2{\max_{x^\prime,z^\prime}{|\langle x^\prime\ket{z^\prime}|^2}}$ randomness. Therefore, the total randomness obtained is $-N\log_2{\max_{x^\prime,z^\prime}{|\langle x^\prime\ket{z^\prime}|^2}}-NH(X')$ and each state generates $-\log_2{\max_{x^\prime,z^\prime}{|\langle x^\prime\ket{z^\prime}|^2}}-H(X')$ randomness on average. In practice, the dephasing and distillation process can be equivalently achieved with the recently proposed coherence distillation protocols, which can be further reduced to a randomness extraction procedure.
Therefore, the final randomness output for two general imperfect bases is
\begin{equation}\label{3}
  R_1 = -2n_Z\log_2{\max_{x^\prime,z^\prime}{|\langle x^\prime\ket{z^\prime}|}}-n_ZH(e_{bX}+\theta)-t_e.
\end{equation}
Note that the randomness output only depends on the term $\max_{x^\prime,z^\prime}{|\langle x^\prime\ket{z^\prime}|}$ instead of a full description of the $X^\prime$ and $Z^\prime$ bases.

In addition, we also consider the case that the measurement efficiencies in the two eigenstates are different. Suppose the efficiencies of projecting onto $\ket{0}$ and $\ket{1}$ are given by $\eta_0$ and $\eta_1$, respectively. Then according to the standard analysis in QKD\cite{Fung:Mismatch:2009},  the randomness output will be further modified to
\begin{equation}\label{4}
  R_{\textrm{final}} = \frac{2\min(\eta_0,\eta_1)}{\eta_0+\eta_1}[-2n_Z\log_2{\max_{x^\prime,z^\prime}{|\langle x^\prime\ket{z^\prime}|}}-n_ZH(e_{bX}+\theta)-t_e],
\end{equation}
That is, the total randomness is rescaled with a factor $\frac{2\min(\eta_0,\eta_1)}{\eta_0+\eta_1}\le1$. The maximum value of 1 can be achieved when $\eta_0 = \eta_1$. Here, we assume that the adversary has no information of the detection efficiency mismatch, otherwise there may exist attacks in analogy to the time-shift attack from QKD\cite{qi2005time}. With such an assumption, the factor can be understood as a simple strategy that we randomly discard the measurement outcome of the higher efficiency detector such that the effective efficiencies of the two detectors are the same.

In experiment, the term ${\max_{x,z}{|\langle x\ket{z}|}}$ and the efficiency $\eta_0,\eta_1$ can be first measured during a calibration procedure on the measurement device. Then, the SI-QRNG scheme can be applied to produce randomness according to the randomness rate formulas in Eq. \eqref{4}.

\subsection{Experimental realization}

\begin{figure}[htbp]
\centering\includegraphics[width=9cm]{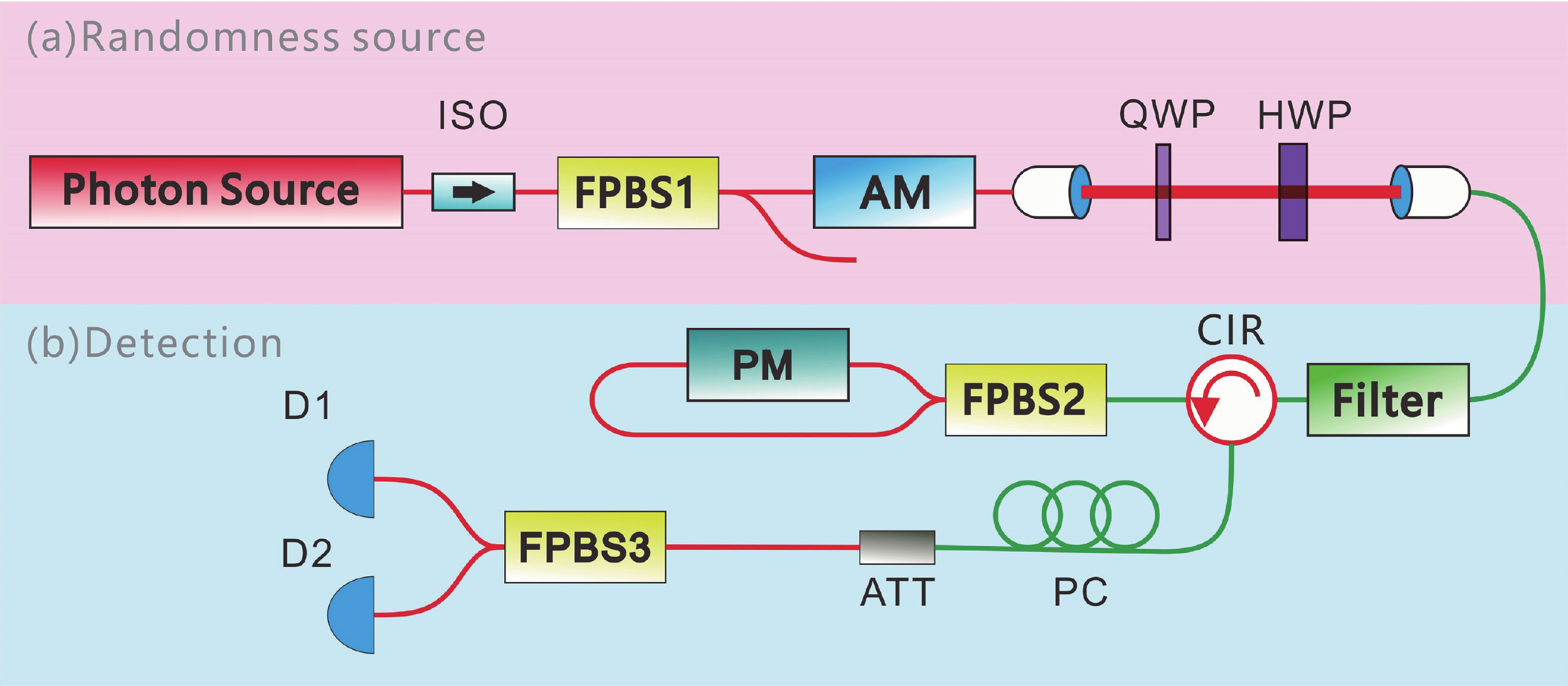}
\caption{The setup of the experiment can be divided into the photon source and detection parts. In the detection part, a Sagnac type interferometer with a phase modulator (PM) is used to select the measurement basis by applying a controllable unitary operator. After proper attenuation, the photons are detected by two up-conversion single photon detectors. In the randomness source part, a photon source is modulated by an amplitude modulator (AM) and transmit through a polarized beam splitter (PBS), a half-wave plate (HWP) and a quarter-wave plate (QWP) to prepare the desired state. The photon source used here is a cw laser and the sunlight, and can be replaced by any other light if necessary. ISO: optical isolator; CIR: optical circulator; FPBS: fibre polarized beam splitter; ATT: attenuator; PC: polarization controller.}
\label{fig:setup}
\end{figure}

As shown in Fig.~\ref{fig:setup}, the experiment setup can be accordingly divided into two parts, the randomness source part and the detection part. While the detection part should be elaborately designed and carefully calibrated, the randomness source part can be uncharacterized or even untrusted.

In the detection part, filters in several dimensions are employed to rule out unexpected optical modes. Spectral filters, including two 100 GHz DWDMs and several interference filters, are used to guarantee that only photons with expected wavelength can enter, and the isolation on unwanted wavelength is over 60 dB. The coupling of single mode fibre excluded unwanted spatial modes. Finally, photons arrived at wrong time will be inspected and eliminated by a time-digital converter (TDC). The selection of measurement basis is realized by a Sagnac type interferometer and a phase modulator (PM) to obtain high visibility and stability, as shown in Fig.~\ref{fig:setup}. For an input pulse with arbitrary polarization state of $\alpha\ket{H}+\beta\ket{V}$, where $\alpha^2 + \beta^2 = 1$, it is split by a fibre polarized beam splitter (FPBS2) when entering the Sagnac interferometer. The length of fibre from PM to one port of FPBS2 is 25.2 meter shorter than to the other port. Thus, the time for the clockwise ($\ket{H}$) and anti-clockwise ($\ket{V}$) parts of the split pulses reach the PM are separated by around 126 ns, and finally back to FPBS2 at the same time. By carefully control the PM, the two parts of pulse can be applied by different phase, named $\varphi_c$ and $\varphi_a$. After combined again in FPBS2, the state of output pulse is $\alpha e^{i\varphi_c}\ket{H}+\beta e^{i\varphi_a}\ket{V}$, correspond to a unitary operation of $U_{F}$.
 A fibre polarization controller is employed to perform an additional unitary operation of $U_{C}$. Here,
\begin{equation}\label{2U}
U_{F}=\left(\begin{array}{ccc}
e^{i\varphi_c} & 0  \\
0 & e^{i\varphi_a}
\end{array}
\right), ~~~~~~~~~ U_{C}=\frac{1}{2} \left(\begin{array}{ccc}
1+i & 1-i  \\
1-i & 1+i
\end{array}
\right)
\end{equation}
Finally, after appropriate attenuation, the pulse is separated by FPBS3 and detected by two up-conversion single photon detectors with efficiency of $10\%$, dark count of 200 cps and dead time of 50 ns.\cite{Shentu:UpConvertion:2013} In this way, we can choose to perform $Z$ ($X$) basis measurement by setting $\varphi_c$ to be  0 ($-\pi/4$) and $\varphi_a$ to be 0 ($\pi/4$). The probability of measuring in the $Z$ ($X$) basis is selected as $99.6\%$ ($0.4\%$) in our experiment and the average photon number per pulse is selected around $13.9$ before detection to maximize the generation rate of random number. Generally, higher photon number per pulse brings higher error rate in the $X$ basis and higher double clicks probability in the $Z$ basis that need to be discarded, while lower photon number per pulse leads to lower $n_z$. Thus, there is a tradeoff for choosing a proper average photon number. The details of optimizing the probability of measuring in the $Z$ ($X$) basis and the average photon number per pulse is discussed in Methods.

As aforementioned, the detection may have imperfections. Therefore, the detection part is first calibrated by an auxiliary cw laser diode with expected wavelength of 1550.12 nm. Considering the imperfect measurement basis of $X^\prime$ and $Z^\prime$, an additional process is performed to estimate $\max_{x^\prime,z^\prime}{|\langle x^\prime\ket{z^\prime}|}$ in \eqref{4}. Firstly, the input state is prepared as the eigenstate of $Z^\prime$ basis, that is, the ratio of photon counting between detector 1 and detector 2 is above 30 dB under $Z^\prime$ basis measurement. Then, in the $X^\prime$ basis measurement, the ratio of photon counting between detector 1 and detector 2 is measured and the bound of $-2\log{\max_{x^\prime,z^\prime}{|\langle x^\prime\ket{z^\prime}|}}$ is calculated to be 0.952.

Although the randomness source part can be untrusted, to demonstrate the high generation rate of the setup, a carefully calibrated randomness source is realized. An amplitude modulator (AM) is used to modulate the input photons to pulses with frequency of 4 MHz and the full width at half maximum (FWHM) of 100 ns. Another fibre polarized beam splitter (FPBS1), a half-wave plate (HWP) and a quarter-wave plate (QWP) is used to prepare the desired polarization state for the detection part.

As the photon source can be any light that does not need to be trustable, the choice of photon source is flexible. Here, we also demonstrate the use of the most common light in the nature --- the sunlight, as the photon source to generate random numbers. A collimator mounted on an equatorial mount is placed on the rooftop to collect sunlight into a single mode fibre. The sun can be approximately considered as a area light source with divergence angle around 0.5$^\circ$.\cite{Fontani:2013dd} Thus, a common collimator with focus length of 11 mm is enough to collect sufficient photon intensities. About 49 nW of light can be collected into single mode fibre under a good weather after filtered by a 1550$\pm$1.5 nm bandpass filter.

\begin{figure}[htbp]
\centering\includegraphics[width=8cm]{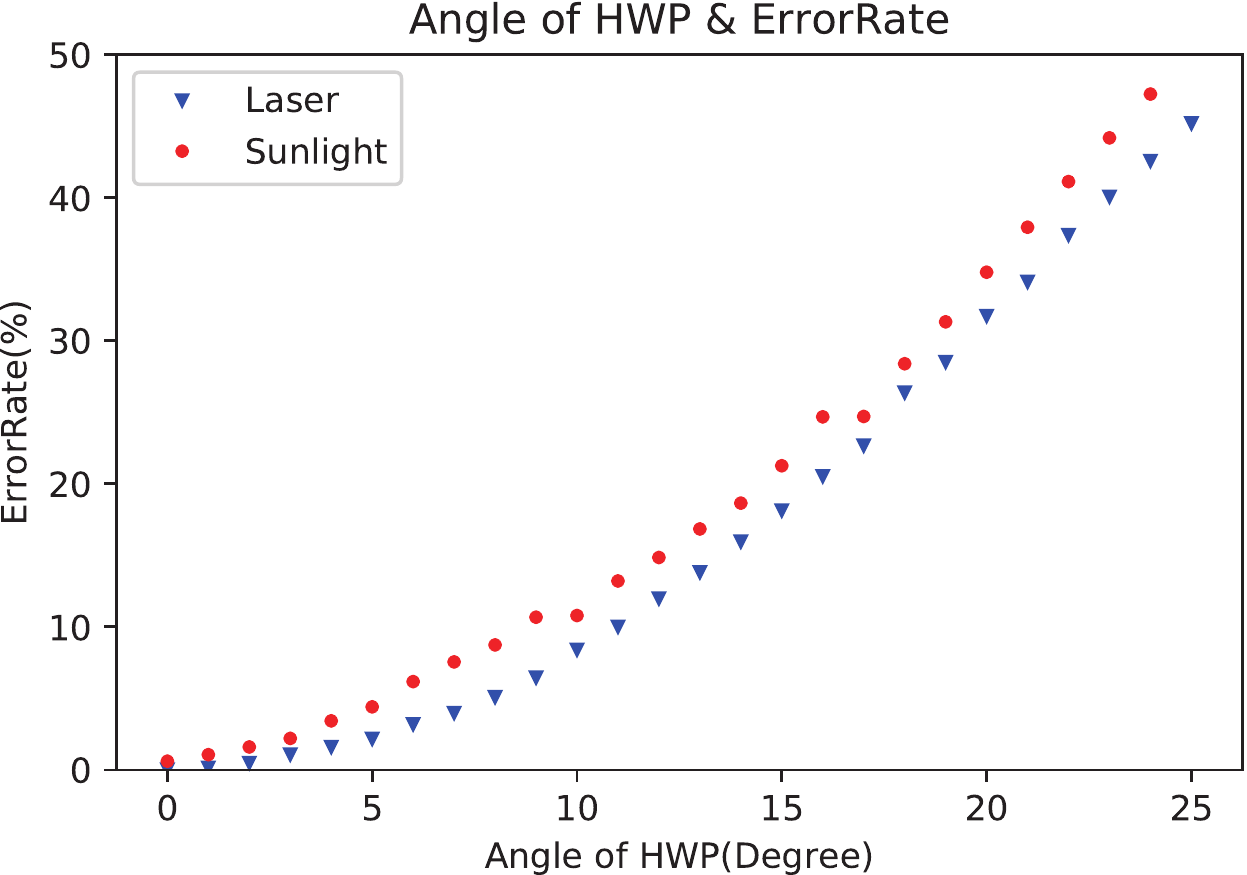}
\caption{For our experiment, when the input state is $\ket{+}$, the bit error rate close to zero. Rotating the HWP change the input state while the bit error rate. This Figure shows the relationship between the angle of HWP and the error rate.}
\label{fig:ber}
\end{figure}

\begin{figure}[htbp]
\centering\includegraphics[width=10cm]{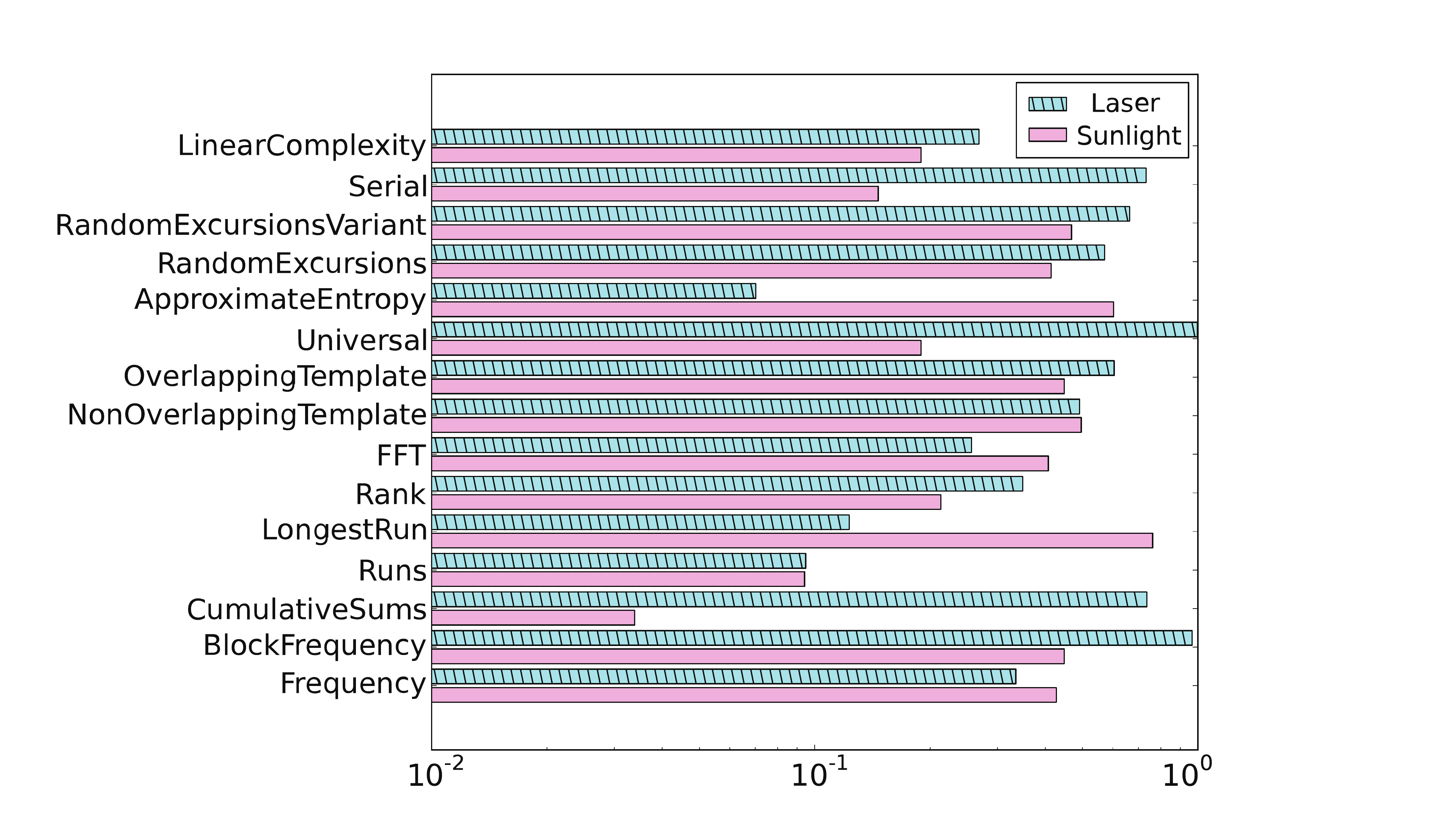}
\caption{The P value of NIST tests. Blue represents laser and red represents sunlight.}
\label{fig:nist}
\end{figure}

The optimal input state for the detection part is the eigenstate of $X$ basis. However, input state with other polarization state does not affect the reliability of randomness. Although the error rate in the $X$ basis will increase and the random number generation rate will reduce. By rotating the HWP in randomness source part to different angles, the relationship between the input state and the error rate of $X$ basis measurement is shown as Fig. \ref{fig:ber}. Under the near optimal condition with the input state of $|+\rangle$, we performed the experiment for both laser and sunlight. The error rate in the $X$ basis measurement is $0.33\%$ for laser and $0.21\%$ for sunlight. The quantum random number generation rate is 1.81 Mbps for laser and 1.72 Mbps for sunlight. The detailed results is shown as Table \ref{tab:result}. The extracted random bits can pass NIST randomness test as shown in Fig.~\ref{fig:nist}.

\begin{table}[htbp]
\centering
\caption{\bf Experiment Result}
\begin{tabular}{ccc}
\hline
  Photon source & Laser & Sunlight\\
\hline
APD1 click at Z basis (count) & 1733623848
 & 1638255301
\\
\hline
APD2 click at Z basis (count) & 1843484418
 & 1725825404
\\
\hline
Bit error rate at X basis (\%) & 0.33 & 0.21\\
\hline
Time (s) & 1800 & 1800 \\
\hline
$\theta$ & 0.001 & 0.001\\
\hline
Extracted raw number (bit)  & $3.26 \times 10^9$ & $3.10 \times 10^9$\\
\hline
Generation rate (bps)  & $1.81 \times 10^6$ & $1.72 \times 10^6$\\
\hline
\end{tabular}
  \label{tab:result}
\end{table}

\section*{Discussion}
In this work, we theoretically modified the SI-QRNG scheme by considering practical issues of measurement devices and experimentally demonstrated the applicability of the scheme in generating reliable and fast random numbers. Compared to the proof of principle demonstration in the theoretical work\cite{Cao16} whose randomness generation rate is about $10^{-3}$ Mbps, our implementation improved the generation speed over 3 orders. Therefore, our SI-QRNG scheme can be applied in many scenarios where both the randomness generation speed and reliability are required. The randomness generation rate here is mainly limited by the detection rate of the single photon detector. Improving the detection rate of the single photon detector can thus further increase the randomness generation rate.

Our result highlighted the power of the state-of-the-art quantum technology. In previous works, it was shown that randomness can be obtained by measuring the intensity of an LED light with a mobile phone\cite{Mobile2014} or by measuring the arrival time of photons from cosmic sources\cite{Zeilinger:CosmicBelltest:prl2017,Qiang:Cosmic:arxiv2017}. However, such QRNG schemes are based on physical models of the LED light or the cosmic source. In our work, we showed that such assumptions are not necessary. Even with a common light in the nature --- sunlight, we can still generate randomness both reliably and fast. Since no assumption is made on the photon source, the coherence or photon number statistics of the photon source does not affect the randomness of the extracted bits. In future works, by exploiting the SI-QRNG scheme, it is also interesting to modify (in theory) and realize (in experiment) those QRNG schemes such that the assumption of the source is removed.

\section*{Methods}
\subsection{Optimizing the generation rate}
\label{opt}

To optimize the final quantum random number generation rate, some proper parameters should be chosen or measured in Eq. (\ref{4}). The first term $\frac{2\min(\eta_0,\eta_1)}{\eta_0+\eta_1}$ depends on the efficiencies of the two detectors, and has a maximal value of 1 when the two efficiencies are equal. Thus, $\eta_0$ and $\eta_1$ are configured to be approximatively uniform ($\eta=10\%$) in our experiment. The second term $-2\log_2{\max_{x^\prime,z^\prime}{|\langle x^\prime\ket{z^\prime}|}}$ depends on the accuracy of controlling the PC and the PM. Due to the imperfection of the actual measurement basis, this term is calculated to be 0.952. $t_e$ is chosen as 100.\cite{PhysRevA.87.062327} Other terms are related to the average photon number $\lambda$ per pulse before detection. A low average photon number lowers $n_z$, while a high average photon number brings higher $e_{pz}$ and higher double click probability in the Z basis. We can rewrite the final random number generation rate as following
\begin{equation}\
  r_{\textrm{final}} =G \cdot p_{z(single-click)}(0.952-H(e_{bX}+\theta)) - 100.
\end{equation}
Here, G stands for the repetition rate of squash speed and equal to 4 MHz in our experiment. $p_{z(single-click)}$ is the probability that one and only one detector clicks for a pulse. As Poisson distribution for laser and multi-mode sunlight,
\begin{equation}\
p_{z(single-click)} = 2e^{-\frac{\lambda^\prime}{2}}(1-e^{-\frac{\lambda^\prime}{2}}),
\end{equation}
where $\lambda^\prime = \lambda \cdot \eta$. For large $n_X$, $e_{pZ}$ and $e_{bX}$ can be regarded as the same. The relationship between $\lambda$ and the final raw rate is shown int Fig. \ref{fig:ber}. The optimal $\lambda$ is about 14.4. For sunlight, the actual $\lambda$ of 11.6 is slightly deviated from the optimal value, due to the intensity fluctuate of sunlight. However, for a wide range of $\lambda$ the random number generation rate is not obviously dropped.

\begin{figure}[hbt]\centering
\resizebox{8cm}{!}{\includegraphics{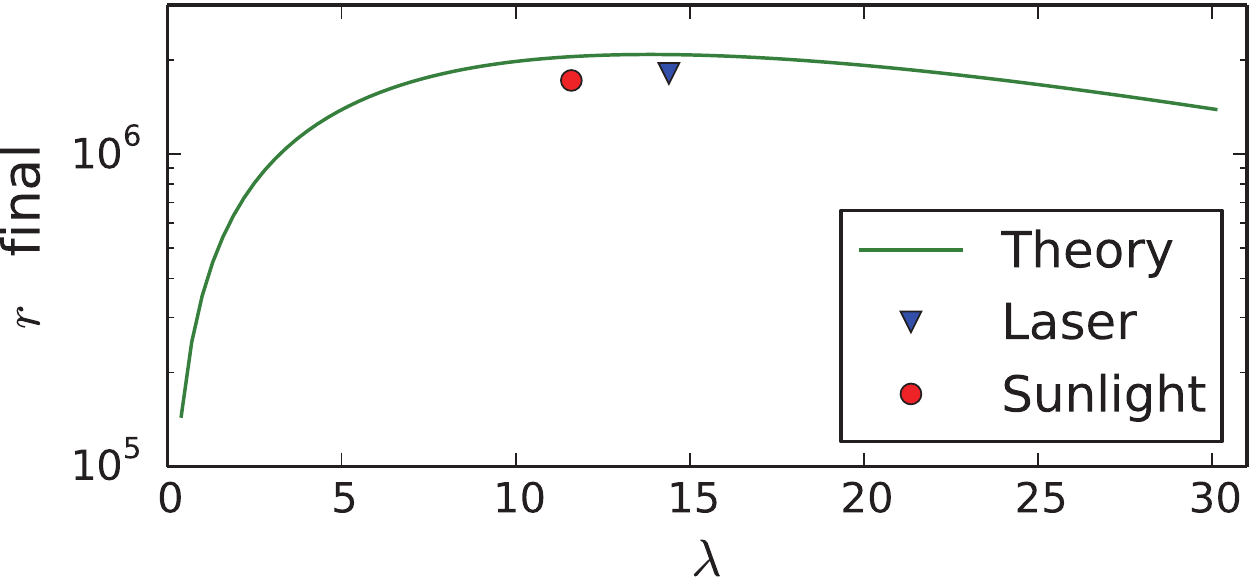}}
\caption{Relationship between the average photon number $\lambda$ per pulse and final random number generation rate. $r$ have the maximal value for $\lambda = 13.9$. In this experiment, $\lambda$ is $14.4$ for laser and is $11.6$ for sunlight. The circle and the triangle denote the actual key generation rate for the two photon sources, respectively. These two values are slightly deviated from the theoretical curve owing to the imperfections.}
\label{fig:rate}
\end{figure}

\begin{addendum}
 \item[Data Availability] Data available on request from authors.
 \item[Author Contributions] Y.-H.L., X.H., Y.C. and J.Y. designed the experiment; Y.-H.L., X.H., Z.-P.L. and J.-Y.G. performed research; Y.-H.L., X.H., and X.Y. analyzed data; X.Y. and X.M. provided the theoretical support; Q.Z., C.-Z.P. and J.-W.P. supervised the project. All authors contributed to the research and the preparation of the manuscript.
 \item This work was supported by the National Natural Science Foundation of China (under Grant No. 11654005, U1738201, U1738142), the National Key Research and Development Program of China (under Grant No. 2017YFA0303901), the Shanghai Sailing Program, and the Chinese Academy of Sciences (CAS). Yuan Cao was supported by the Youth Innovation Promotion Association of CAS (under Grant No. 2018492). This is a post-peer-review, pre-copyedit version of an article published in npj Quantum Information. The final authenticated version is available online at: http://dx.doi.org/10.1038/s41534-019-0208-1.
 \item[Competing Interests] The authors declare that there are no competing interests.
 \item[Correspondence] Correspondence and requests for materials
should be addressed to Cao Yuan~(yuancao@ustc.edu.cn), Xiongfeng Ma~(xma@tsinghua.edu.cn) or Cheng-Zhi Peng~(pcz@ustc.edu.cn).

\end{addendum}

\bibliographystyle{naturemag}

\newpage

\end{document}